\documentclass[twocolumn,showpacs,preprintnumbers,amsmath,amssymb]{revtex4}

\usepackage{graphicx}
\usepackage{dcolumn}
\usepackage{bm}

\begin{document}

\preprint{APS/123-QED}

\title{Ferroelectric ordering and electroclinic effect\\in chiral smectic liquid crystals}

\author{Yu. G. Fokin}
\email{yura@shg.ru} \homepage{http://www.shg.ru}
\author{T. V. Murzina}
\author{O. A. Aktsipetrov}
 \affiliation {Physics Department, Moscow
State University, Moscow, 119992, Russia}

\author{S. Soria}
\altaffiliation[At present at]{ ICFO-Institut de Ciencies
Fotoniques, and Dept of Signal Theory and Communications,
Universitat Politecnica de Catalunya,  08034 Barcelona, Spain}

\author{G. Marowsky}
\affiliation{ Laser-Laboratorium Goettingen e.V.,
Hans-Adolf-Krebs-Weg 1, D-37077 Goettingen, Germany }

\date{\today}

\begin{abstract}
Ferroelectric ordering, the electroclinic effect and chiral
smectic C (SmC*) - smectic A (SmA*) phase transitions in thin
planar ferroelectric liquid crystal (FLC) cells are studied by
means of linear electrooptic and second harmonic generation
techniques. The ferroelectric switching is detected in biased FLC
cells by measuring azimuthal dependences of linear and nonlinear
responses.  The applied DC-electric field rotates the FLC symmetry
axis with initial and final orientations in the cell plane.
Comparative studies of the switching behavior in reflection and
transmission allows to distinguish the contributions from the bulk
and the sub-surface layers of the cell. The analysis of
temperature dependence shows the existence of a strong surface
coupling. The temperature dependent nonlinear polarization shows a
critical behavior corresponding to the superfluid model.
\end{abstract}

\pacs{64.70.Md, 42.65.Ky, 77.80.Fm, 61.30.Hn }

\maketitle
\section{\label{sec:level1}Introduction }

\par Ferroelectric liquid crystals (FLC) have been studied intensively for several
decades. Chiral smectic liquid crystals (LC) have unique material
properties like spontaneous polarization ~\cite{M}. Although a
single chiral smectic LC molecule has a non-zero dipole moment due
to symmetry considerations ~\cite{M1}, LC molecules in the bulk of
the sample tend to form a helical structure which leads to
polarization compensation. The helix can be unwound by
application of a strong electric field or using thin test cells.
Conventionally, alignment of LC is usually obtained by
unidirectional mechanical rubbing of a thin polymer layer which
coats the inner cell surfaces. Thus, the LC molecules point their
long  axes along the rubbing direction. This anisotropic
interaction influences the ferroelectric ordering and the
switching behavior which are of great practical importance for any
LC device. The ferroelectrically ordered SmC* phase is
characterized by a non zero angle between the molecular long axis
and the smectic layer normal. This tilt angle is the order
parameter used to describe the second order SmC*-SmA* phase
transition.

\par In several studies the temperature dependence of the tilt angle
below the critical temperature follows a power law of 0.5
(classical behavior) ~\cite{M4}, whereas in other studies it
follows a power law of 0.3, in strong analogy with the superfluid
helium predicted by de Gennes ~\cite{M5}. The SmA* phase in the
vicinity of the transition point has also been the subject of
intensive investigations, since in that part of the critical
region the electroclinic (EC) effect can induce a tilt angle which
is proportional to the electric field strength ~\cite{M7}. The
induced tilt angle is also strongly temperature dependent and
diverges as the phase transition temperature is approached from
above. Another type of this phenomenon, the so-called surface EC
effect, is observed in the interfacial region of the FLC cells and
originates from the interaction of the sub-surface layers of
chiral smectic molecules and a localized surface field
~\cite{ShenXue}. In the vicinity of the transition point, the
critical exponents of the tilt angle are in the range of 0.5
$\div$ 1.5 ~\cite{ShenXue, M5, M7, Selinger}. The surface EC
effect results in the existence of  sub-surface region in which
the director is twisted from the rubbing axis to the bulk
alignment direction. The ferroelectric properties of these twisted
layers are of great importance as they play a dominant role in
LC-devices.

\par In this paper we use linear electrooptic (EO) and
second harmonic generation (SHG) techniques for investigation of
ferroelectric ordering, SmC*-SmA* phase transitions and the EC
effect in thin FLC cells. The main emphasis is made on comparative
studies of the SHG reflection and transmission experimental
geometries in order to figure out the roles of the sub-surface
layers and the bulk in ferroelectric phase transitions. The SHG
method is well-known for its sensitivity to symmetrical,
structural and electronic properties of surfaces, interfaces and
ultrathin films, and has been widely used along with EO for
studying ferroelectric properties of chiral smectic
LC~\cite{shgeo}. As the SHG response strongly depends on the polar
state of the matter because of its unique sensitivity to the
breakdown of the inversion symmetry ~\cite{M13}, it is a powerful
instrument for probing ferroelectric phase transitions and
electric-field induced effects in the vicinity of the critical
region.

\section{Model description}
\begin{figure}
\includegraphics[width=0.8\columnwidth]{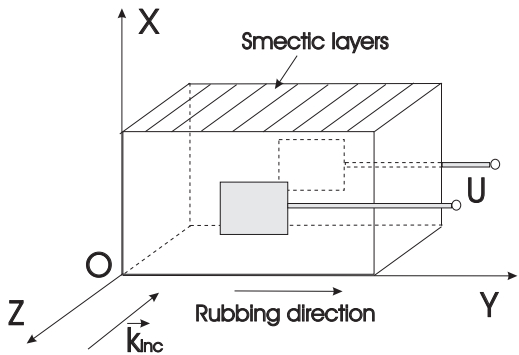}
\caption{\label{geom} Schematic representation of experimental
geometry.}
\end{figure}

We studied linear and nonlinear quadratic responses of the FLC
cells upon application of an electric field and temperature
variation up to the SmC*-SmA* phase transition. Chiral smectic LC
belong to the $C_{2}$ point group symmetry with the symmetry axes
oriented in the cell plane. A schematic representation of the FLC
cell and the coordinate system is shown in fig.~\ref{geom}. In
this coordinate frame $Oz$ is the normal to the cell, $XY$ is the
cell plane and the $C_{2}$ axis is parallel to the $Oy$ direction
and coincide with the main optical axis. We assume the FLC cell to
be uniaxial, with the ordinary and extraordinary refractive
indices $n_{0}=n_{x}=n_{z}$, and $n_{e}=n_{y}$, respectively. The
transmittance of an optically active plate of the thickness $d$
and birefringence ${\Delta}n$ at the wavelength $\lambda$ in
crossed-polarizers geometry is determined by the equation:
\begin{equation}
T=\sin^{2}(2\alpha)\sin^{2}\frac{{\Delta}n{\pi}d}{\lambda},
\label{lintran}
\end{equation}
where $\alpha$ is the angle between the main optical axes and
polarization direction of the incident light. The linear
electro-optic tensor $r_{ijk}$, where indices \textit{i, j, k}
denote the axes of the cell coordinate system, for the given
symmetry has 8 non-vanishing components ${r}_{xxy}$, ${r}_{yyy}$,
${r}_{zzy}$, ${r}_{yzx}$, ${r}_{yzz}$, ${r}_{xzy}$, ${r}_{xyx}$,
${r}_{xyz}$ ~\cite{M16}. Application of the DC electric field
\textit{E} along $Oz$ direction leads to the rotation of the
optical axes, which could be interpreted in terms of the rotation
of the ellipse of refractive indexes. The equation of index
ellipsoid in the presence of the electric field $E$ has the form
\begin{equation}
\frac{x^{2}}{n^{2}_{x}}+\frac{y^{2}}{n^{2}_{y}}+\frac{z^{2}}{n^{2}_{z}}+2yz{r}_{yzz}E+2xy{r}_{xyz}E=1.
\label{ellipsfield}
\end{equation}
The transformation of (\ref{ellipsfield}) to the canonical form
gives the DC field-induced rotation of the undisturbed system of
coordinates based on the main optical axes. Considering the
rotation only around $Oz$-axis, the influence of the electric
field can be described by the tilt of the main optical axis in the
cell plane with the dependence of the turn angle ${\Delta}\alpha$
on the electric field as:
\begin{equation}
\Delta\alpha=\frac{1}{2}\arctan\frac{2r_{xyz}E}{\frac{1}{n_{o}^{2}}+\frac{1}{n_{e}^{2}}}.
\label{angle}
\end{equation}
The nonlinear quadratic susceptibility tensor of the FLC cell with
a $C_{2}$ symmetry is given by following non-vanishing components
${\chi}_{ijk}$, ${\chi}_{yxx}$, ${\chi}_{yyy}$, ${\chi}_{yzz}$,
,${\chi}_{yzx}$, ${\chi}_{xyz}$, ${\chi}_{xxy}$, ${\chi}_{zyz}$,
${\chi}_{zxy}$ ~\cite{M17}. The contribution to the SHG intensity
of each quadratic susceptibility component depends on the angle of
incidence of the fundamental radiation, on the azimuthal position
of the sample and polarizations of the input and SHG light. In
transmission geometry at normal incidence only three components
participate in SHG - ${\chi}_{yxx}$, ${\chi}_{xxy}$ and
${\chi}_{yyy}$, and for \textit{pp} -geometry the SHG intensity
can be expressed by:
\begin{equation}
{I}_{pp}^{2\omega}\sim\cos^{2}\theta({\chi}_{yyy}\cos^{2}\theta +
(2{\chi}_{xxy}+{\chi}_{yxx})\sin^{2}\theta)^{2}, \label{pp}
\end{equation}
where the azimuthal angle $\theta$ is the angle between the
direction of the \textit{p}-polarization and ${C}_{2}$ axes of the
cell symmetry. By fitting the anisotropy dependences of the SHG
intensity in all polarization combinations, the corresponding
components of the quadratic susceptibility can be extracted.

\par According to a well-known model, in the smectic planar layers FLC
molecules can precess on the surface of the smectic cone
~\cite{cones18}. Ferroelectric switching is attributed to the
interaction between molecular dipoles and external electric field
and results in rotation of the molecules and, correspondingly, of
the symmetry axes, within the half of the smectic cone. If the
molecular director positions, corresponding to the saturating
electric fields of  opposite values, are in the cell plane and on
the opposite sides of the smectic cone, then (\ref{angle}) can be
used for the explanation of linear and nonlinear switching, as it
shows the in-plane rotation of the main optical axis.

The angle between the molecular long axis and the normal to the
smectic layers, i.e. the apex angle of the cone, is the order
parameter of the SmC*-SmA* phase transition, vanishing in the SmA*
phase. It can also have a finite magnitude in the SmA* phase and
diverge in the vicinity of the critical temperature due to the EC
effect. The temperature-induced changes of this angle lead to the
changes in $\alpha$ and $\theta$ angles which are determined by
the orientation of the main axis. We show below that the
temperature and electric field dependences of the main axis
orientation can be deduced from linear and nonlinear optical
experiments, in order to describe the ferroelectric switching,
phase transitions and the EC effect in FLC cells.

\section{Experiment}
The principle arrangement of the experimental setup is described
elsewhere ~\cite{M15}. Briefly, an OPO laser system is used  as a
source of the fundamental radiation, with the output wavelength of
537 nm, repetition rate 10 Hz and pulse duration 4 ns. An
appropriate set of color filters is used for linear and nonlinear
effects measurements. A photodiode or a PMT tube and gated
electronics are used as a registration system for linear or
nonlinear experiments. The temperature is varied from 20 to 55
${^{o}C}$, and automatically controlled by a digital thermocouple
thermometer with the accuracy of ${1^{o}C}$. An electric field in
the range of -15 to +15 MV/m is applied along $OZ$ axis through
the ITO electrodes. Commercial FLC cells (E.H.C. Co., Tokyo) of a
nominal thickness of 2 ${\mu}m$ are used.  These cells have an
unidirectional rubbed polyimide surface layer. The samples are
prepared by a slow cooling of the mixture ~\cite{M15}, capillary
filled in the isotropic phase. The critical temperature of the
SmC*-SmA* phase transition for the studied mixture is about
${42^{o}C}$.

\par
\begin{figure}
\includegraphics[width=0.8\columnwidth]{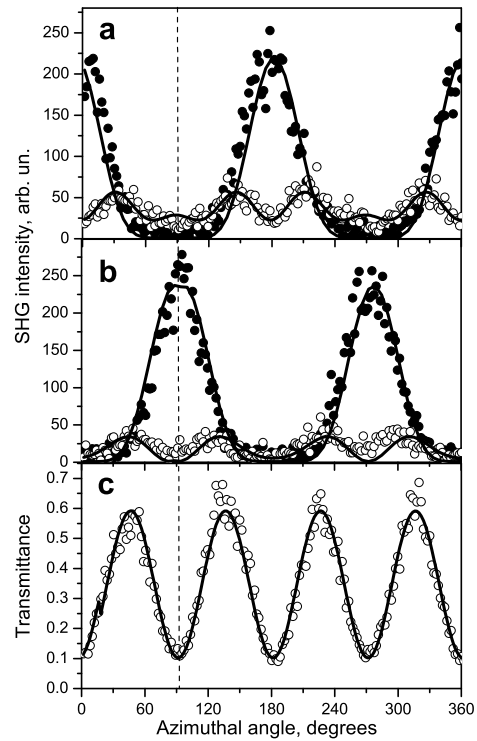}
\caption{\label{anisall} Anisotropy dependence of the SHG
intensity in transmission, a. \textit{ss} (filled circles) and
\textit{sp} (open circles) and b. \textit{pp} (filled circles) and
\textit{ps} (open circles) geometries; c. Anisotropy dependence of
the linear transmittance at cross polarizers geometry. Solid lines
are calculated from theory.}
\end{figure}

 To determine the in-plane components of the nonlinear
susceptibility, the SHG intensity vs. azimuthal angle is measured
at normal incidence in transmission geometry for different
polarization combinations - \textit{pp}, \textit{ss}, \textit{ps},
\textit{sp}, where the first letter denotes polarization of the
incident light, and the second one - polarization of the light
transmitted through the analyzer, placed after the sample. The
observed anisotropy is shown in fig.~\ref{anisall}a, b and
correlates well with the dependence (\ref{pp}), obtained for a
${C_{2}}$ symmetry structure. No electric field is applied.
\par Ferroelectric switching of the cells is studied by measuring the
anisotropy dependences of the SHG intensity at different values of
the external DC electric field applied along the normal to the
cell plane. Fig.~\ref{nelsw}a shows the SHG intensity in
transmission geometry for \textit{pp}-polarizations and for two
applied DC fields of opposite sign. The dependences are shifted
relative to each other, the phase difference is about 30$^{o}$. In
fig.~\ref{nelsw}b the dependences of the SHG intensities vs.
electric field are shown for different anisotropy positions of the
cell. The anisotropy position determines the contrast value of the
switching dependences and the ratio between SHG responses
corresponding to opposite biases. Switching in reflection geometry
also exhibits a shift of the anisotropy dependences for opposite
electric fields with a phase difference of about 20$^{o}$.

\par Linear transmittance anisotropy is measured in crossed polarizers
geometry and is shown in fig.~\ref{anisall}c. Application of the
electric field leads to a shift of the linear transmittance
anisotropy, as in the nonlinear case. In both cases, we observed a
shift of about 30$^{o}$.

\par Temperature dependences of the SHG response are taken in
reflection geometry at incident angle about 45$^{o}$ and in
transmission at normal incidence. The cases of the presence and
absence of external DC field are studied. Fig.~\ref{figtemp}a
shows temperature dependences of the SHG intensity in \textit{pp}
transmission geometry for biased (filled circles and squares) and
unbiased (open circles) cells. In the absence of the electric
field the temperature dependence of the SHG intensity approaches a
constant value at critical temperature in accordance with a power
law at $T<T_{c}$, and shows no manifestation of the surface EC
effect at $T>T_{c}$ in the SmA* phase. For biased LC samples, a
critical behavior of the SHG intensity due to the EC effect is
obtained in the SmA* phase in the vicinity of the phase
transition, which is seen from the divergent behavior of the SHG
intensity at $T>T_{c}$, similar to the critical behavior of the
tilt angle.

\section{Discussion}

\par FLC cell structure usually contains layers with different direction of
spontaneous polarization - twisted or helicoidally wounded layers.
This modulation of the space orientation of molecular dipoles is
governed by the intermolecular forces and substrate influence. We
suppose that in the absence of the DC electric field the cell has
a net dipole moment and axis of $C_{2}$ symmetry laying in the
cell plane. Then it is possible to approximate anisotropy
dependences of SHG intensity in all possible polarization
combinations in transmission at normal incidence by formulas
analogous to (\ref{pp}) for \textit{pp} case. In
 fig.~\ref{anisall} calculated curves are shown for all geometries.
The approximation gives the ratios between in-plane components of
nonlinear susceptibility:
\begin{equation}
{\chi}_{xxy}:{\chi}_{yyy}:{\chi}_{yxx}=1:23.9:-3.3 \label{inplane}
\end{equation}

\par The linear transmittance anisotropy has four equal peaks and no
dark extinction, which is typical for twisted structures. Assuming
the main optical axis to lay in the cell plane, approximation of
the linear anisotropy dependence by (\ref{lintran}) gives a
birefringence of  $\sim0.041$, which corresponds to the
characteristic values for smectic LC measured by other authors
~\cite{shgeo}. Comparison of nonlinear and linear anisotropy
dependences indicates that ${C}_{2}$ symmetry axis, corresponding
to the maximum of the SHG anisotropy in \textit{pp}-geometry, and
the main optical axis, corresponding to the minimum of the linear
transmittance anisotropy in \textit{sp}-geometry, are parallel.
\begin{figure}
\includegraphics[width=0.8\columnwidth]{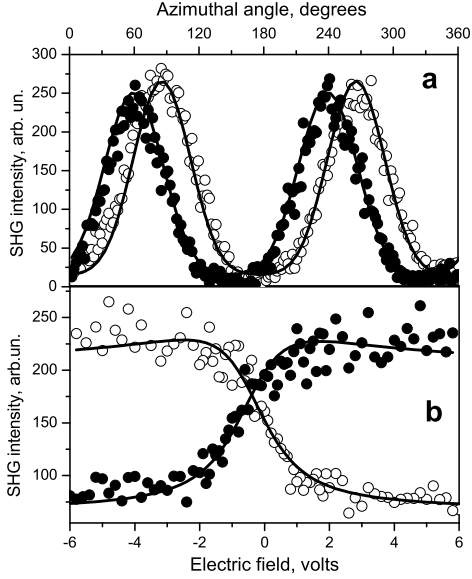}
\caption{\label{nelsw}
 a. Anisotropy dependence of the SHG
intensity in transmission geometry at different electric field
applied: +8V (filled circles), -8V (open circles). Solid lines are
calculated by (\ref{anisfield}); b. SHG intensity dependence on
the electric field in transmission for anisotropy positions 100
(open circles) and 230 (filled circles). Solid lines are
calculated by (\ref{anisfield}) and (\ref{arctg}). }
\end{figure}

\par Temperature dependences of the SHG intensity for
the applied electric field of different polarities reveal a
hysteresis free SmC*-SmA* second order phase transition. A typical
critical behavior in the vicinity of the SmC*-SmA* phase
transition is obtained. A non-zero SHG signal in the SmA* phase
denotes that some part of the cell is still oriented and have a
polar order with an in-plane component of the nonlinear
polarization. The temperature dependence of the SHG response for
reflection geometry shows qualitatively the same behavior as in
transmission geometry, which is connected with the SmC*-SmA* phase
transition in the sub-surface layer of the cell. The SHG
temperature dependence can be explained by the interference of the
field and temperature dependent and independent contributions to
nonlinear polarization, responsible for the SHG signal. Then, we
can write for the nonlinear polarization
\begin{equation}
{\bf \overrightarrow{P}}_{NL}^{2\omega}=
{\bf\overrightarrow{P}}_{\textit{surf}}^{2\omega}(E,T)+{\bf
\overrightarrow{P}}_{\textit{bulk}}^{2\omega}(E,T)+{\bf
\overrightarrow{P}}_{\textit{const}}^{2\omega}, \label{temp}
\end{equation}
where ${\bf\overrightarrow{P}}_{\textit{surf}}^{2\omega}(E,T)$ and
${\bf\overrightarrow{P}}_{\textit{bulk}}^{2\omega}(E,T)$ are
electric field $E$ and temperature $T$ dependent contributions to
the nonlinear polarization at double frequency from the
sub-surface layers and the bulk, respectively, and
${\bf\overrightarrow{P}}_{\textit{const}}^{2\omega}$ is a field
and temperature independent component. For the transmitted SHG the
correlation length is about 5  microns, while for reflection it is
about 0.1 micron~\cite{M12}, which indicates that in reflection a
thin sub-surface layer participates in SHG. As SHG temperature
dependences of sub-surface layers and the bulk have the same
qualitative character, then either
${\overrightarrow{P}}_{\textit{bulk}}^{2\omega}(E,T)$ and
${\overrightarrow{P}}_{\textit{surf}}^{2\omega}(E,T)$ from
(\ref{temp}) should be the same functions of $E$ and $T$, or the
bulk contribution to nonlinear polarization should vanish.

\par Explanation of temperature dependences demands the existence of
an electric field and temperature independent in-plane
contribution ${\overrightarrow{P}}_{\textit{const}}^{2\omega}$ to
the nonlinear polarization. Orientation of the molecules in the
bulk layers of the cell can be electric-field independent in some
experimental geometries depending on the layer packing, but
temperature dependence is obligatory because of the presence of
the SmC*-SmA* phase transition. Then the non-switching layer
should be in the sub-surface region. We can assume that a "frozen"
sub-surface layer, with a thickness smaller than the correlation
length for SHG in reflection geometry, exists. It is strongly
stabilized by the surface coupling, and does not respond to any
external electric field or temperature variations, as well as the
layer closest to the substrate. The latter has only
\textit{z}-component of the nonlinear polarization, as the dipole
moments of its molecules are directed along the normal to the
substrate plane, and then it does not contribute to the SHG at
normal incidence. But the rest of the "frozen" region contains
layers with in-plane components of spontaneous polarization due to
the twist near the surface, and thus participates in the SHG at
any angle of incidence.

\par
\begin{figure}
\includegraphics[width=0.8\columnwidth]{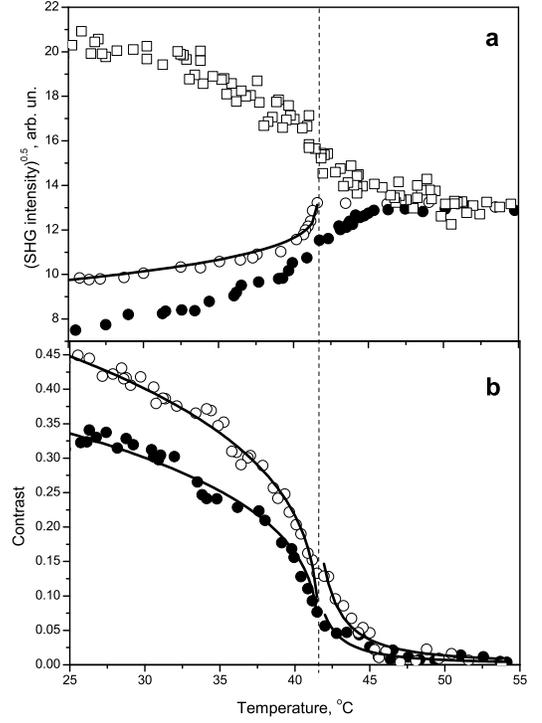}
\caption{\label{figtemp} a. Temperature dependence of the SHG
intensity in transmission geometry at different applied electric
field : +8V (squares), 0V (open circles), -8V (filled circles).
Solid line is a fit by (\ref{TE0}); b. Temperature dependence of
the contrast in reflection (filled circles) and transmission (open
circles) geometries. Solid lines are calculated by (\ref{Kup}) and
(\ref{Kdown}). }
\end{figure}

 Let us figure out the interconnection between the temperature
dependences of SHG and the order parameter of the SmC*-SmA* phase
transition. Usually the order parameter of this phase transition
is the tilt angle, as spontaneous polarization is compensated
because of the helical structure. The thickness of the cells
studied in this paper is smaller than the helix pitch, besides,
anisotropic interaction with the substrate produces an additional
orientational order of the FLC molecules, so that the cells have
spontaneous polarization ${\overrightarrow{P}_{\textit{sp}}}$ in
SmC* phase. ${\overrightarrow{P}_{\textit{sp}}}$ decreases in SmA*
phase and is a linear function of the tilt angle ${\theta(E,T)}$.
So we can write for the SHG intensity:
\begin{equation}
\sqrt{I^{2\omega}}\sim {\overrightarrow{P}}_{NL}^{2\omega}\sim
{\overrightarrow{P}}_{\textit{sp}}\sim\theta(E,T),
\label{orderpar}
\end{equation}
assuming that ${\sqrt{I^{2\omega}(E,T)}}$ has the same critical
dependence on temperature as the tilt angle. Then, according to
(\ref{temp}) and (\ref{orderpar}), the temperature dependent
contribution ${\sqrt{I^{2\omega}(E,T)}}$  in the absence of the
electric field and for $T<T_{c}$ can be approximated by the
following expression:
\begin{equation}
\sqrt{I^{2\omega}(E=0)}\sim
P_{const}^{2\omega}-P_{0}^{2\omega}(1-\frac{T}{T_{c}})^{\beta},
\label{TE0}
\end{equation}
where $P_{const}^{2\omega}$ and $P_{0}^{2\omega}$ are proportional
to isotropic and anisotropic contributions of nonlinear
polarizations, $T_{c}$ is the temperature of the phase transition,
and $\beta$ is the critical exponent. Approximation of the curve
is shown in fig.~\ref{figtemp}a by a solid line, with
$\beta=0.31$, in correspondence to the theory of phase transition
in a superfluid helium.

\par To compare the temperature behavior of the nonlinear polarization
in the vicinity of the phase transition for reflection and
transmission cases, we introduce the contrast of the dependences
as
\begin{equation}
K=
\frac{\sqrt{I^{2\omega}(+E)}-\sqrt{I^{2\omega}(-E)}}{\sqrt{I^{2\omega}(+E)}+\sqrt{I^{2\omega}(-E)}},
\label{contrast}
\end{equation}
where $I^{2\omega}(\pm{E})$ are the SHG intensities for positive
and negative voltages applied to the cell. In fig.~\ref{figtemp}b
the contrast vs. temperature is shown for reflection and
transmission geometries. The contrast determined in such a way
shows a behavior similar to that  of the order parameter near the
phase transition. Introducing SHG intensity dependences on
temperature in the presence of the electric field at $T<T_{c}$ in
the way, analogous to (\ref{TE0}),
\begin{equation}
\sqrt{I^{2\omega}({\pm}E)}\sim
P_{const}^{2\omega}{\pm}P_{0}^{2\omega}({\pm}E)(1-\frac{T}{T_{c}})^{\beta},
\label{TE}
\end{equation}
and supposing $P_{0}^{2\omega}(E)=P_{0}^{2\omega}(-E)$, we can
express the contrast \textit{K} as:
\begin{equation}
K\sim\frac{P_{0}^{2\omega}(E)}{P_{const}^{2\omega}}(1-\frac{T}{T_{c}})^{\beta}.
\label{Kup}
\end{equation}
Approximation of experimental dependences by (\ref{Kup}) in the
SmC* phase gives $\beta=0.31$ (fig.~\ref{figtemp}b), as in the
case of the absence of the field. The amplitude of the contrast
(\ref{Kup}) depends on the ratio
$P_{0}^{2\omega}(E)/P_{const}^{2\omega}$ of field dependent and
independent contributions to the nonlinear polarization. In
transmission this ratio can be bigger due to the larger amount of
layers participating in ferroelectric switching, besides, the
field independent contribution of the boundary layer leads to the
contrast decrease in reflection geometry. These considerations
explain why the contrast of temperature dependences and electric
field-induced shift of anisotropy dependences in transmission are
bigger than in reflection geometry, as obtained in the experiment.
\par Temperature dependence of the contrast in the region of the EC
effect, at $T>T_{c}$, can be written in the same way, considering
the temperature dependence  ${\sqrt{I^{2\omega}(E,T)}}$ similar to
the temperature dependence of the tilt angle above the critical
temperature:
\begin{equation}
\sqrt{I^{2\omega}({\pm}E)}\sim
P_{const}^{2\omega}{\pm}P_{0}^{2\omega}({\pm}E)(1-\frac{T}{T_{c}})^{-\gamma},
\label{EC}
\end{equation}
 where  $\gamma$  is the critical exponent,
 which gives under the same considerations the contrast for
 $T>T_{c}$:
\begin{equation}
K\sim\frac{P_{0}^{2\omega}(E)}{P_{const}^{2\omega}}(1-\frac{T}{T_{c}})^{-\gamma},
 \label{Kdown}
\end{equation}
Approximation of experimental data (fig.~\ref{figtemp}b) gives
$\gamma=1.4$. Thus we observe that both critical exponents are in
agreement with the superfluid helium model.

\par Application of a DC electric field to the FLC leads to a shift of
anisotropy dependences of the linear and nonlinear responses, as
shown in fig.~\ref{nelsw}a. Similar pictures are obtained in the
reflection case. As the character of anisotropy dependences is the
same in the presence of the field, our suggestion of the switching
mechanism based on the symmetry axis rotation in the cell plane
seems to be rather realistic. The shift originates from the
rotation of the molecules in the switching layers, the resulting
nonlinear polarization of the cell is the sum of unchanged
quadratic polarization from the "frozen" layer and from
field-dependent region with the new position of the symmetry axes.
Thus for approximating the SHG anisotropy dependence in the
presence of the electric field, we divide the nonlinear
polarization into two parts, and (\ref{pp}) can be rewritten as:
\begin{equation}
{I}_{pp}^{2\omega}\sim(gP_{NL,pp}^{2\omega}(\theta+d\theta)
+(1-g)P_{NL,pp}^{2\omega}(\theta))^{2}, \label{anisfield}
\end{equation}
where $g$ is the effective thickness of switching layer relatively
to the non-switching one and $d\theta$ is a field-induced change
of the azimuthal angle. The value of $g$ is estimated to be 0.5
which indicates that the magnitude of the field-independent
contribution is comparable to the field-dependent one, allowing
effective mutual interference in accordance with (\ref{temp}).
Symmetry axis shift $d\theta$ between opposite values of the
voltage is about 52$^{o}$. For electric field dependences of the
SHG intensity, shown in fig.~\ref{nelsw}b, the independent
variable is the electric field and in (\ref{anisfield}) the
azimuthal angle $\theta$ is fixed by the anisotropy position. Then
$d\theta$ becomes a function of the electric field, chosen,
according to (\ref{angle}), as a saturation function:
\begin{equation}
{d\theta(E)=b\arctan(c(E-dE))}, \label{arctg}
\end{equation}
where $b$ and $c$ are constants, depending on the strength of
interaction between field and angle, and $dE$ is a constant
determined by the switching history, as ferroelectric switching of
SHG intensity exhibits characteristic field hysteresis.
Experimental dependences in fig.~\ref{nelsw}b are taken for two
anisotropy positions with the difference in ${\theta}\sim130$
degrees. Fitting one of the curves by (\ref{anisfield}) with
(\ref{arctg}) allows to extract $b=0.5$ and $c=0.63$. Varying then
only $dE$ and ${\theta}$ during approximation of the second curve
gives the difference of the anisotropy position of about 126
degrees, in good agreement with experimental conditions. Values of
$dE$ in both cases lay in the region of typical hysteresis width
and do not exceed 1 V/micron. The value of $b$ corresponds to the
tilt of the symmetry axis in the cell plane, discussed above
(\ref{angle}).

\par In conclusion, the electroclinic effect, the SmC*-SmA* phase
transition and ferroelectric switching have been studied in thin
planar cells of ferroelectric chiral liquid crystals by means of
electrooptic and second harmonic generation techniques. The
analysis of the temperature dependences of nonlinear quadratic
response in the critical region leads to the assumption of a
strong surface coupling existence, resulting in the stabilizing of
several "frozen" subsurface twisted layers, independent on
electric field and temperature. Critical exponents in the vicinity
of the SmC*-SmA* phase transition show a behavior corresponding to
the superfluid helium theory. Ferroelectric switching has been
observed in linear and nonlinear anisotropic responses and has
been explained by the model, assuming in-plane rotation of
symmetry and main optical axes of the cells in the presence of
external electric field.
\\
\\
\par
\begin{acknowledgments}
We thank Prof. G. Heppke's group for synthesis of the samples used
in our work. The work was supported by INTAS grant 2002-113/F1b,
President grant "Leading Russian Scientific Schools" 1604.2003.2
and EU network grant ERB FMRXCT 980209 (SILC). S.S acknowledges
funding from European Regional Development Fund and from the
Ministerio de Ciencia y Tecnologia through the Ramon y Cajal
program.
\end{acknowledgments}



\end{document}